\documentclass[a4paper,11pt]{article}
\usepackage[T2A]{fontenc}
\usepackage[utf8]{inputenc}
\usepackage{url}
\def\lesssim{\mathrel{\hbox{\rlap{\hbox{\lower4pt\hbox{$\sim$}}}\hbox{$<$}}}}

\def\gtrsim{\mathrel{\hbox{\rlap{\hbox{\lower4pt\hbox{$\sim$}}}\hbox{$>$}}}}

\makeatletter
\def\@xfootnote[#1]{%
  \protected@xdef\@thefnmark{#1}%
  \@footnotemark\@footnotetext}
\makeatother

\title{Exploding neutron stars in close binaries}
\author{S. I. Blinnikov, I. D. Novikov, T. V. Perevodchikova, and A. G. Polnarev}
\date{}

\begin{document}

\maketitle

\begin{abstract}
The discovery of GW signal from merging neutron stars by LIGO on 17th August 2017 was followed
by a short GRB170817A discovered by FERMI and INTEGRAL 1.7 seconds after the loss of the GW signal when it just reached its maximum.
Here we present a reproduction of the first paper (published by us in 1984) predicting a short GRB after GW signal of merging neutron stars.
Our paper followed the scenario by Clark and Eardley (1977) who predicted a catastrophic
disruption of a neutron star in a binary 1.7 seconds after the peak of GW signal.
Our next paper in 1990 predicted all the main properties of the short GRB with quite a reasonable accuracy.
Typos in English translation are corrected and a few comments are added in the current publication as numbered footnotes (the only footnote from  the original paper is marked by an asterisk).
\end{abstract}

\newpage

\begin{center}

{\Large {\bf Exploding neutron stars in close binaries}
\bigskip

S. I. Blinnikov, I. D. Novikov, T. V. Perevodchikova, and A. G. Polnarev}

\bigskip

\textit{Institute of Theoretical and Experimental Physics, Moscow \\
and Institute for Space Research, USSR Academy of Sciences, Moscow}

(Submitted January 27, 1984)

Russian original: Pis'ma Astron. Zh. \textbf{10}, 422--428 (June 1984)

English translation: Sov.Astron.Lett. \textbf{10}, 177-179
(May-June 1984)\footnote{\url{http://adsabs.harvard.edu/abs/1984SvAL...10..177B
}}

\end{center}

\begin{small}

A close binary system comprising a neutron star and another neutron star (or a black hole) will evolve so that
the less massive component sheds mass, passing through a series of quasiequilibrium states, until it achieves
its minimum possible mass $m_{\min} \approx 0.09 M_\odot$ and explodes.
In a compact globular cluster or the nucleus of a
galaxy, such evolution can terminate in an explosion in less than the Hubble time.

\end{small}

\bigskip

Several authors have pointed out~[1 -- 3] 
that a neutron
star may eventually explode if its mass drops to the minimum possible equilibrium value,  $m_{\min} \approx 0.09 M_\odot$, due
to nucleon decay.
(At lower values of the mass, no equilibrium states exist with a density comparable to the
nuclear density, and a neutron star would begin to expand.)
Such a process could, of course, occur only in the remote
future, after an interval of the order of the proton decay
time.

Another way for a neutron star to lose its equilibrium
has been mentioned by Page~[2] (and attributed to M.~J.~Rees);
it rests on the hypothesis that the gravitational constant may
be diminishing with time.
Furthermore, a neutron star
may come to be destroyed tidally if it is paired with a
black hole~[4,5] or with another neutron star~[6].

In this letter we wish to point out that the evolution
of a binary system comprising two neutron stars (or a
neutron star and a black hole, although for brevity we
henceforth omit special mention of this case) will, after
the stage of tidal mass loss (tidal disruption, in Clark and
Eardley's terminology [6]\footnote{More accurate term used in [6] is ``tidal
stripping'' followed by the ``tidal disruption''.}), result in one of the neutron
stars losing its hydrostatic equilibrium as its mass falls
below $m_{\min}$, so that it will explode. Neutron star
explosions of this kind could therefore be taking place today
and can be handled by conventional astrophysics without
resorting to exotic hypotheses.

One may envisage the following general scenario for
the processes leading to the explosion of a neutron star.
A close binary system will evolve into a pair of gravitationally bound neutron stars.
If this evolution takes place
in a reasonably compact globular cluster or in a star
cluster at the center of a galaxy, then the two components
may gradually approach each other, relinquishing energy
as they interact with other stars in the cluster.
When the
binary becomes tight enough its evolution will be dictated
by the emission of gravitational waves.
Eventually the
less massive, larger-radius neutron star will fill its entire Roche lobe,
and matter will begin to flow toward its
more massive companion.
If certain criteria are satisfied
[see inequality (8) below], the mass transfer will occur
at a pace determined by the rate at which gravitational
waves are radiated by the system.
When the smaller
star has achieved a certain critical mass, a substantially
more rapid transfer will set in due to tidal forces. These
two stages have been described by Clark and Eardley~[6];
we shall indicate some refinements that are needed.
Finally the mass of the lighter neutron star will drop below $m_{\min}$,
producing hydrodynamic instability and an
explosion. We proceed now to consider certain aspects
of this scenario more fully.

First let us estimate how much time the binary system requires to evolve to the point
where one of the neutron stars overfills its Roche lobe.
After both of the binary components have become neutron stars,
the evolution of the pair will initially be controlled by interaction
with individual cluster members passing nearby.
If at the outset the pair has a binding energy greater in absolute value
than the mean kinetic energy of a star in the
cluster (a ``hard'', or ``tight'', binary), then gravitational
interactions with passing stars will on the average serve
to increase the binding energy, and the binary components will approach each other.

To make some very simple estimates, suppose that
the two neutron stars and the cluster members all have
the same mass $m$. The tight-pair condition may be written
(see, for example, Dokuchaev and Ozernoi~[7]) as
\begin{equation}
{Gm^2 \over 2a} > {mv^2 \over 3} ,
\end{equation}
where $a$ is the major semiaxis of the binary orbit (assumed circular) at starting time,
and $v$ is the velocity dispersion of the stars in the cluster.
Thus a tight binary will have a radius less than
\begin{equation}
 a_1 = 15 R_\odot \left(\frac{m}{10^{33}\,\mbox{g}} \right) \left( \frac{v}{100\,\mbox{km/sec}} \right)^{-2} .
\end{equation}
The evolution time to a state with an orbit of radius $a$
can be estimated from the expression
\begin{equation}
 t \approx (10^{10}\,\mbox{yr}) \left( \frac{v}{100\,\mbox{km/sec}}\right) \left(\frac{a}{R_\odot}\right)^{-1} \left(\frac{m}{10^{33}\,\mbox{g}} \right)^{-1} \left(\frac{n}{10^{7}\,\mbox{pc}^{-3}} \right)^{-1}  ,
\end{equation}
where $n$ is the number density of cluster members and
$R_\odot$ is the radius of the sun. Dokuchaev and Ozernoi
estimate~[7] that gravitational radiation will begin to dominate
the approach rate of the binary components when $a$ has
diminished to the point where
\begin{equation}
 \frac{a_1}{a} \approx 10 \left(\frac{m}{10^{33}\,\mbox{g}} \right)^{3/5} \left(\frac{n}{10^{7}\,\mbox{pc}^{-3}} \right)^{1/5} \left( \frac{v}{100\,\mbox{km/sec}}\right)^{-11/5} .
\end{equation}
Denote this value of $a$ by $a_2$. From Eqs. (2), (4) we have
\begin{equation}
 a_2  \approx {R_\odot} \left(\frac{n}{10^{7}\,\mbox{pc}^{-3}} \right)^{-1/5}
  \left( \frac{v}{100\,\mbox{km/sec}}\right)^{1/5}
 \left(\frac{m}{10^{33}\,\mbox{g}} \right)^{2/5}  .
\end{equation}
On substituting the value (5) into Eq.~(3) we find that the
evolution time until $a_2$ is reached will be
\begin{equation}
 t \approx (10^{10}\,\mbox{yr}) \left( \frac{v}{100\,\mbox{km/sec}}\right)^{4/5} \left(\frac{n}{10^{7}\,\mbox{pc}^{-3}} \right)^{-4/5}  \left(\frac{m}{10^{33}\,\mbox{g}} \right)^{-7/5}   .
\end{equation}

Gravitational-wave losses will cause the components
to approach each other on a time scale
\begin{equation}
 t_{\rm grav} \approx (10^{10}\,\mbox{yr}) \left(\frac{m_1}{10^{33}\,\mbox{g}} \right)^{-1}
 \left(\frac{m_2}{10^{33}\,\mbox{g}} \right)^{-1}
 \left(\frac{m_1}{10^{33}\,\mbox{g}} + \frac{m_2}{10^{33}\,\mbox{g}}\right)^{-1} \left(\frac{a}{R_\odot}\right)^4 ,
\end{equation}
where $m_1, \; m_2$ are the masses of the components\footnote{Important analytical refinements to the
formulas for gravitational-wave losses, and the whole scenario, see in
Imshennik, V.~S., \& Popov, D.~V.\ 1994, Astronomy Letters, 20, 529;
Imshennik, V.~S., \& Popov, D.~V.\ 1998, Astronomy Letters, 24, 206;
Imshennik, V.~S., \& Popov, D.~V.\ 2002, Astronomy Letters, 28, 465, and references therein.}.
Since
the gravitational-radiation rate increases as $a$ becomes
smaller, the total time required for the binary to evolve
from the state with $a \approx a_1$~[Eq.~(2)] to an arbitrarily
close pairing will be given by the estimate (6).

From the expression (7) we may infer that over the
life of the Galaxy, neutron star binaries will indeed have
been able to tighten so greatly that the less massive,
larger component (henceforth subscript 2) will overfill
its Roche lobe. How will the binary evolve next?

To begin with, we would point out that the crystalline
structure of neutron star surface layers will not prevent
the transfer of matter from the overflowing star to its
companion, because at the Lagrangian point the material
will be weightless, the pressure will drop, and the crystals will melt.
Moreover, at such distances the tidal
forces will be enormous, breaking down the crystals.
Nevertheless, the crystalline envelope may affect the rate
at which matter reaches the Lagrangian point, thereby
altering the mass-transfer rate.
Analysis of the evolution of a close binary dominated
by gravitational-wave emission~[8-10] shows that if the total
mass of the system remains constant and the criterion
\begin{equation}
 \frac{d\ln R_2}{d\ln m_2} > 2 \frac{m_2}{m_1} - {5 \over 3}
\end{equation}
is satisfied ($R_2$ is the radius of the star with $m = m_2$;
$m_2 < m_1$), then the rate of evolution will be determined
by the gravitational-wave energy losses.
The corresponding mass-loss rate is given by the expression (assuming
the time scale for transfer of accretion-disk angular momentum
to the orbital motion is not longer than the characteristic time for the mass loss)
\begin{equation}
 \dot m_2 = 10^{-9} \frac{m_1 m_2 (m_1+m_2)}{a^3 \left[\frac{2a (m_2-m_1)}{m_1 m_2}- {da \over dm_2}\right] } M_\odot/\mbox{yr} ,
\end{equation}
where $m_1, \; m_2$ are the component masses in solar units
and $a$ is the separation of the components in units of $R_\odot$
The quantity $a$ is related to $R_2$ (taken as the radius of the
Roche lobe) and $m_1, \; m_2$ by
\begin{equation}
a = R_2/[0.462 (m_2/(m_1 + m_2))^{1/3}];
\end{equation}
tables of neutron star structure~[12] provide the $R_2(m_2)$
relation.

If the inequality (8) is violated, a regime of tidal
mass exchange will set in.
For a mass ratio $m_2/m_1$ appreciably below $0.5$,
the limiting condition (8) will be
reached for neutron stars of mass $m_2 = m_{\rm cr} \approx 0.15 M_\odot$.

Accordingly let us give three time scales for a neutron star binary with components of mass
$m_2 = m_{\rm cr} = 0.15 M_\odot, \; m_1 = 1 M_\odot$:
a) the time $t_{\rm gr,cr}$ required for the
parameters to change due to gravitational radiation;
b) the hydrodynamic time $t_{\rm hyd,cr}$, determined by the mean
density $3m_2/4\pi R^3$~[3] of the star;
c) the orbital period $P_{\rm cr}$
at the instant when the inequality (8) is reversed.
Taking
$a_{\rm cr} \approx 78$~km, $R_{2,\rm cr} \approx 18$~km, we have
\begin{equation}
t_{\rm gr, cr}\approx 40\,\mbox{sec}, \quad t_{\rm hyd, cr} \approx 10^{-4}\,\mbox{sec},
\quad P_{\rm cr} \approx 1.2\cdot 10^{-3}\, \mbox{sec} .
\end{equation}
We now show that the time scale for the tidal mass-exchange
regime will evidently be much longer than the
hydrodynamic time, so that for a certain interval the star
will be in quasiequilibrium.

Let $R_{\rm Roche}$ denote the polar radius of the Roche
lobe. An expanding star in equilibrium will have
$R_2 > R_{\rm Roche}$.
Suppose that $R_2 < R_{L_2}$ , where $R_{L_2}$ is the distance
from the center of the star to the second Lagrangian point,
and assume that $h \equiv R_2 - R_{\rm Roche} \ll R_2$.
Then the star will lose mass on a time scale~[11]
\begin{equation}
t_{\rm tid} \approx \frac{m_2}{\dot m_2} \approx t_g \left(\frac{R_2}{h}\right)^{1.5+\frac{1}{\gamma -1}} \gg t_g ,
\end{equation}
where $t_g = R/\langle{v_s}\rangle$,
with $\langle{v_s}\rangle$ denoting the mean sound
speed in a neutron star having approximately a polytropic
mass distribution with index $\gamma$.
Here the quantities $R_2$
and $R_{\rm Roche} = R_2 - h$ are both known functions of $m_2$ (this
is true for $R_{\rm Roche}$, since $m_1$ and $a$ are regarded as known
at the time when tidal transfer begins).
Consequently Eq.~(9) gives $\dot m_2$ as a function of $m_2$, and can be integrated out.
We see, however, from the relation (12) itself that the
evolution time will be much longer than $t_{\rm hyd, cr}$ so long as
$h$ is small.

But in due course the premise that $h$ is small will
be violated.
When the mass of neutron star 2 has fallen
to nearly $m_{\min}$, the stellar envelope will abruptly be
stripped away.
At this stage the approximation of a polytrope with a single index $\gamma$ certainly will no longer
describe the mass distribution, and furthermore, the star
will have a radius many times that of its Roche lobe.
For both reasons the arguments given above will cease
to be valid.
Still, star 2 will have only a very small fraction of its mass in its envelope (see below),
so in a crude approximation, to obtain order-of-magnitude estimates,
we may regard the mass loss by star 2 as consisting essentially of the flow
through the Boche lobe, which encloses the great majority of the mass of that star.

Now let us work out the corresponding estimates.
First we find the size of the Roche lobe at the stage when
$m_2$ reaches $m_{\min}$.
Suppose (approximately) that the angular momentum and combined mass of the components are
conserved\footnote[*]{Actually one should allow here for the loss of mass and angular
momentum by the system as matter flows out near the second Lagrangian point.}; then the formula for
the angular momentum implies that
\begin{equation}
 \frac{a''}{a'} = \left( \frac{m'_1 m'_2}{m''_1 m''_2} \right)^2 .
\end{equation}
The prime and double prime designate, respectively, the
time when tidal mass transfer begins and the time when
$m_2 = m_{\min}$.

Taking the particular case $m_1 = 1 M_\odot$ at the start of
tidal transfer, we find that
\begin{equation}
  a'' = 2.3 a'
\end{equation}
and the Roche lobe will expand by a factor
\begin{equation}
 \frac{R''_{\rm Roche}}{R'_{\rm Roche}} = \frac{a''}{a'} \left( \frac{m''_2}{m'_2} \right)^{1/3} .
\end{equation}
Since $R'_{\rm Roche} = R_{2,\rm cr} = 18$~km, Eq. (15) gives
$R''_{\rm Roche} = 34$~km. For the mass of star 2 enclosed within a radius
$r = 34$~km, the ratio of the envelope mass to the mass of
the whole star will be
\begin{equation}
  m_{34}/m_2 \approx 0.01 ,
\end{equation}
confirming our statement that the envelope mass should
be small.

To order of magnitude $ \dot m_2 = -\rho_{34} v_{34} (R''_{\rm Roche})^2$,
where $\rho_{34}, \; v_{34}$ denote the density and sound speed in the
star at distance $r = 34$~km from the center. The neutron
star model~[12] gives
\begin{equation}
 \rho_{34} = 0.7 \cdot 10^{10}\, \mbox{g/cm}^3, \quad v_{34} = 10^9\, \mbox{cm/sec},
  \quad \dot m_2 = - 0.8 \cdot 10^{32}\, \mbox{g/sec} \approx -0.04 M_\odot/\mbox{sec}.
\end{equation}
The process will operate on a time scale
\begin{equation}
 t'' = \frac{m_2}{|\dot m_2|} \approx 4\, \mbox{sec} .
\end{equation}
Estimating the hydrodynamic time for the mass enclosed
within $R''_{\rm Roche}$, we see ($\bar \rho$ denotes the mean density in
the star) that
\begin{equation}
  t_{\rm hyd} \approx \frac{1}{\sqrt{6\pi G \bar\rho}} \approx 0.3\,\mbox{msec} \ll t'' .
\end{equation}
Thus as star 2 approaches the limit $m_{\min}$ it will
lose mass on a time scale much longer than the hydrodynamic time,
and will reach $m_2 = m_{\min}$ through a series
of quasiequilibrium states. Incidentally, if star 2 has a
mass $m_2 \approx m_{\min}$ from the very outset (say $m_2 \approx 0.1 M_\odot$),
then the transfer will begin in the tidal regime and will
take place with $R_{\rm Roche}$ much larger than the estimate
$R''_{\rm Roche} \approx 34$~km, at a rate far slower than indicated by
the time scale (18).

Once having achieved $m_2 = m_{\min}$, star 2 will lose
its hydrostatic stability and will begin to expand at a rate
determined by $t_{\rm hyd}$ and the amended equation of state.
Clark and Eardley~[6] estimate that perhaps one neutron
star may undergo tidal disruption every 100~yr within
a 15-Mpc radius; thus the event would not be exceedingly
rare. Not only should a burst of gravitational waves be
produced~[6], but also a powerful electromagnetic flare\footnote{Here the original Russian word
``вспышка'' is translated as ``flare''. However, in Russian both ``гамма-вспышка'' and
``гамма-всплеск'' are equivalent to the English ``gamma-ray burst''. Nevertheless, ``$\gamma$-ray
burster'' appears in the next paragraph of the English translation.}
(most likely x~rays and $\gamma$~rays). Page~[2] believes that the
explosion may attain an energy of supernova scale, but
the problem awaits a detailed analysis. We intend to
consider this process further in a separate paper.\footnote{This was done in
Blinnikov,~S. I., Imshennik, V. S., Nadezhin, D. K., Novikov, I. D.,
Perevodchikova,~T.~V., Polnarev,~A.~G. 1990, ---
\url{http://adsabs.harvard.edu/abs/1990SvA....34..595B},
where short duration of the burst, the numbers found for $E_{\rm iso}$:  $10^{43} < E_{\rm iso} < 10^{47}$~ergs, for its
spectral range 10-100~keV, and maximum velocity of ejecta
$10^{10}$~cm/sec are all consistent with observations of GRB170817A and its kilonova.
The duration of the burst was found a bit shorter than in GRB170817A, and kinetic energy was
overestimated, but we have discussed there which corrections to our simple model are needed
(account of 3D effects for the flow and accurate neutrino radiation).}

We also have omitted discussion here of the physical
processes that will accompany the mass transfer, such as
the stripping from the star of material with nuclei having
excess neutrons; as these nuclei later decay, $\gamma$-ray
burster phenomena might occur (like the processes that
Bisnovatyi-Kogan and Chech\"{e}tkin~[13] have discussed).

We thank G. S. Bisnovatyi-Kogan, V. S. Imshennik,
and D. K. Nad\"{e}zhin for their interest in the subject and
valuable comments.\footnote{Important papers which followed our line of development:
Colpi, M., Shapiro, S.~L., \& Teukolsky, S.~A.\ 1989, ApJ, 339, 318 -- here a concept of `minisupernova'
is introduced which is referred to as `kilonova' now;
Eichler, D., Livio, M., Piran, T., \& Schramm, D.~N.\ 1989, Nature, 340, 126.
Independent early papers on neutron star mergers as sources of GRBs:
Paczynski, B.\ 1986, ApJL, 308, L43;
Goodman,~J.\ 1986, ApJL, 308, L47;
Goodman,~J., Dar,~A., \& Nussinov, S.\ 1987, ApJL, 314, L7.}

\newpage
\hrule
\bigskip

1. Ya. B. Zel'dovich, foreword to: S. Weinberg, The First Three Minutes,
Basic Books (1977) [Russ. tr., Energy Press, Moscow (1981)].

2. D. N. Page, Phys. Lett. A {\bf 91}, 201 (1982).

3. I. D. Novikov and T. V. Perevodchikova, Preprint Inst. Kosm. Issled.
Akad. Nauk SSSR No. 861 (1984).\footnote{
Astronomicheskii Zhurnal, {\bf 61}, Sept.-Oct. 1984, p. 935-938; Soviet Astronomy, {\bf 28}, Sept.-Oct. 1984, p. 545, 546.}

4. J. M. Lattimer and D. N. Schramm, Astrophys. I. {\bf 192}, L145 (1974).

5. J. M. Lattimer and D. N. Schramm, Astrophys. J. {\bf 210}, 549 (1976).

6. J. P. A. Clark and D. M. Eardley, Astrophys. J. {\bf 215}, 311 (1977).

7. V. I. Dokuchaev and L. M. Ozernol, Piz'ma Astron. Zh. {\bf 7}, 95 (1981) [Sov. Astron. Lett. {\bf 7}, 52 (1981)].

8.  S. C. Vila, Astrophys. J.,  {\bf 168}, 217 (1971).

9. 
  D.~C.~Heggie,\ Binary evolution in stellar dynamics.\ Monthly Notices of the Royal Astronomical Society  {\bf 173}, 729-787 (1975).\footnote{This reference is present in the original Russian text in Pis'ma AZh, but omitted in translation in Sov.Astr.Lett. due to a technical error: ref.8 is followed there by ref.10.}

10. A. V. Tutukov and L. R. Yungel'son, Acta Astron. {\bf 29}, 665 (1979).

11. L. R. Yungel'son and A. G. Masevich, Sov. Sci. Rev. E: Astrophys. Space
Phys. Rev. {\bf 2}, 29, 1983).

12. G. Baym, C. Pethick, and P. Sutherland, Astrophys. J. {\bf 170}, 299 (1971).

13. G. S. Bisnovatyi-Kogan and V.M. Chech\"{e}tkin, Usp. Fiz. Nauk {\bf 127}, 263
(1979) [Sov. Phys. Usp. {\bf 22}, 89 (1980)].

\end{document}